\documentclass[superscriptaddress,amsmath,amssymb,twocolumn,aps,pre]{revtex4}
\usepackage[normal]{subfigure}
\usepackage{dcolumn}% Align table columns on decimal point
\usepackage{bm}% bold math
\usepackage{pxfonts}

\usepackage[normalem]{ulem}
\usepackage{graphicx,color}
\usepackage{subfigure}
\usepackage{bbold}
\usepackage{cancel}
\usepackage[colorlinks,citecolor=red,linkcolor=blue]{hyperref}
\usepackage[usenames,dvipsnames,svgnames]{xcolor}
\usepackage{calligra}
\DeclareMathAlphabet{\mathcalligra}{T1}{calligra}{m}{n}

\newcommand{\ee}{\end{equation}}
\newcommand{\be}{\begin{equation}}

\newmuskip\pFqmuskip

\newcommand*\pFq[6][8]{%
  \begingroup % only local assignments
  \pFqmuskip=#1mu\relax
  \mathcode`\,=\string"8000
  \begingroup\lccode`\~=`\,
  \lowercase{\endgroup\let~}\pFqcomma
  {}_{#2}F_{#3}{\left[\genfrac..{0pt}{}{#4}{#5};#6\right]}%
  \endgroup
}
\newcommand{\pFqcomma}{\mskip\pFqmuskip}

\begin{document}
\title{Surface tension and the Mori-Tanaka theory of non-dilute soft composite solids}

\author{Francesco Mancarella}
\affiliation{Nordic Institute for Theoretical Physics
(NORDITA), SE-106 91 Stockholm, Sweden}
\author{Robert W. Style}
\affiliation{Mathematical Institute, University of Oxford, Oxford OX1 3LB, UK}
\author{John S. Wettlaufer}
\affiliation{Yale University, New Haven, Connecticut 06520, USA}
\affiliation{Mathematical Institute, University of Oxford, Oxford OX1 3LB, UK}
\affiliation{Nordic Institute for Theoretical Physics
(NORDITA), SE-106 91 Stockholm, Sweden}
\begin{abstract}

Eshelby's theory is the foundation of composite mechanics, allowing calculation of the effective elastic moduli of composites from a knowledge of their microstructure.
However it ignores interfacial stress and only applies to very dilute composites -- i.e. where any inclusions are widely spaced apart.
Here, within the framework of the Mori-Tanaka multiphase approximation scheme, we extend Eshelby's theory to treat a composite with interfacial stress in the non-dilute limit.  
In particular we calculate the elastic moduli of composites comprised of a compliant, elastic solid hosting a non-dilute distribution of identical liquid droplets.
The composite stiffness depends strongly on the ratio of the droplet size, $R$, to an elastocapillary lengthscale, $L$.
Interfacial tension substantially impacts the effective elastic moduli of the composite when $R/L\lesssim 100$.
When $R < 3L/2$ ($R=3L/2$) liquid inclusions stiffen (cloak the far-field signature of) the solid.  

\end{abstract}
%\pacs{... ; ... ; ... }
\date{\today}
\maketitle

\section{Introduction}

In a seminal paper Eshelby described the strain response of isolated inclusions to applied stresses, and predicted the stiffness of solid composites containing a dilute volume fraction of inclusions \cite{Eshelby57}.
These results has since been successfully used to model a huge range of problems, from composite mechanics to fracture and dislocation theory.
As Eshelby's theory strictly only applies to composites containing dilute inclusions, it has been extended to treat non-dilute composites with a variety of approximation schemes \cite{Hashin63,Hashin62,Christensen79,Cauvin07,Fornes03}, many of which show good agreement with experimental data across an unexpectedly wide range of inclusion volume fraction $\phi$.

Although Eshelby--theory and its non-dilute extensions work well for hard composites, recent work has shown that they can fail to describe soft composites \cite{Stylesoft15, Style15}.
This is because such schemes view the constituents of the composite as bulk linear-elastic solids, while ignoring the physics of the interface between them \cite{Eshelby57,Hashin63,Mori73,Christensen05,Hashin62,Christensen79}.
However, as is generally the case in interfacial thermodynamics \cite{Cahn78}, when the inclusions become sufficiently small that the surface energy becomes appreciable relative to the bulk strain energy, one cannot ignore interfacial effects.
For example, when the interface between an inclusion and the host (with Young's modulus $E$) is governed by an isotropic, strain-independent surface tension $\gamma$, the validity of the standard framework \cite[e.g.,][]{Eshelby57,Hashin63,Mori73,Christensen05,Hashin62,Christensen79} is limited to inclusions much larger than the elastocapillary length $L\equiv \gamma/E$ \cite{Stylesoft15, Style15}.
This is typically the situation for soft materials such as gels and elastomers \cite[e.g.,][]{Hui13,styl12c,mora13,nade13}.

Here we extend Eshelby's theory to soft, non-dilute composites with an isotropic, strain-independent interfacial surface tension.
In particular, motivated by recent experiments and their analysis \cite{Stylesoft15, Style15}, we focus on the problem of a soft elastic solid containing a non-dilute distribution of identical liquid droplets.
The framework for our extension is the multiphase scheme introduced by Mori and Tanaka \cite{Mori73}, and our approach 
generalizes previous theoretical results that have been compared with experiments on soft aerated composites \cite{Palierne90,Linh13,Ducloue14}.
Our work differs from previous approaches that either consider dilute inclusions or interfacial elasticity \cite[e.g.,][]{Sharma04,Duan05JMPS,LeQuang07,Stylesoft15}, or that obtain upper and lower bounds on composite elastic moduli with interfacial elasticity \cite{LeQuang08,Brisard10,Brisardbulk10}.

\section{The Mori-Tanaka, or Equivalent Inclusion-Average Stress (EIAS), method}\label{approxschemes}

\begin{figure}[htp]
\centering 
\includegraphics[width=0.9\columnwidth]{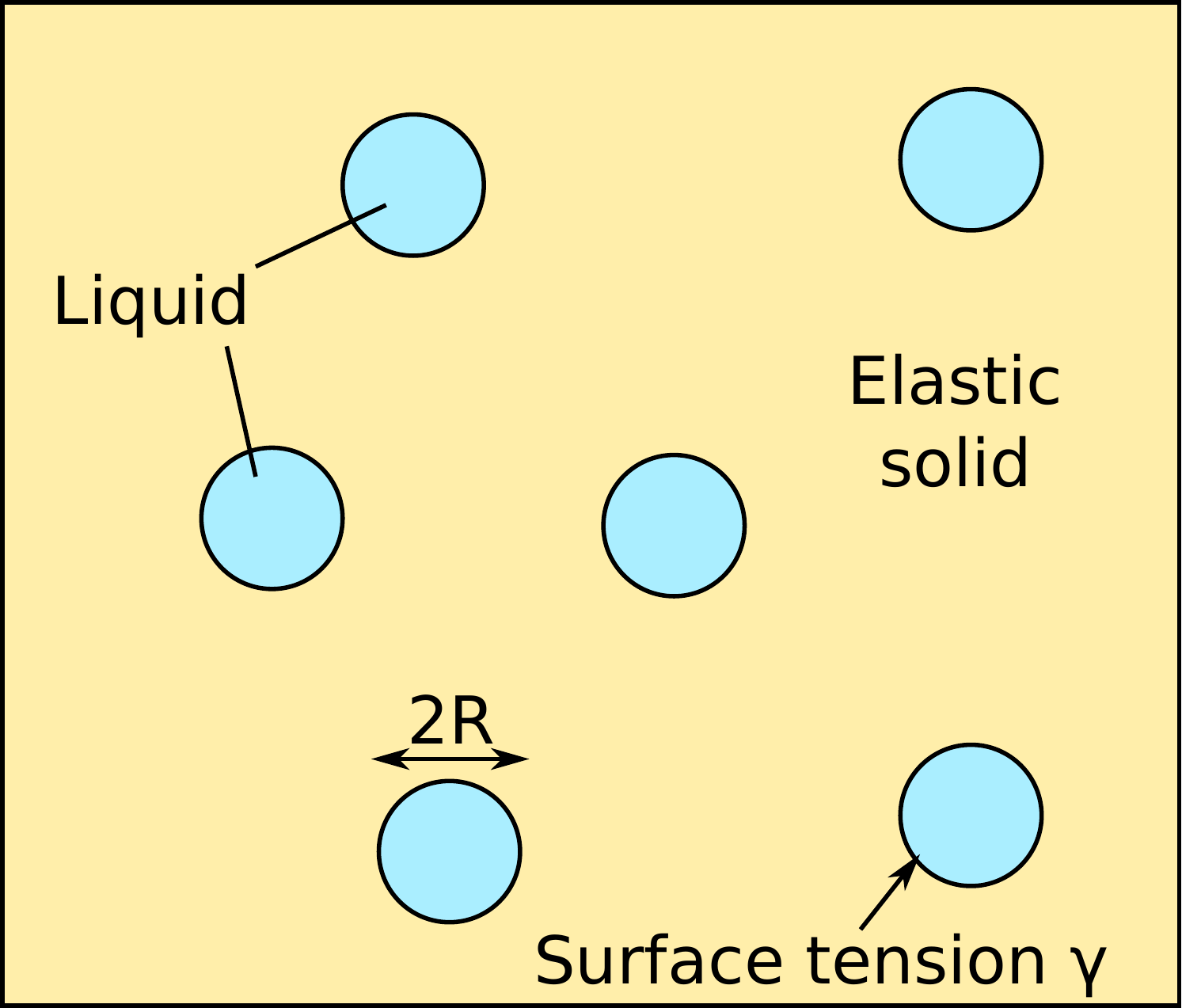}
\caption{Schematic representation of the composite material treated. Identical liquid inclusion droplets are embedded in a solid elastic matrix. }
\label{fig:schematic}
\end{figure}

The concept of an equivalent inclusion \cite{Eshelby57} and the average-stress in the matrix are central to the Mori-Tanaka approximation scheme, which we refer to as Equivalent Inclusion-Average Stress (EIAS) method \cite[e.g.,][]{Benveniste87}.  Here, we envision a two-phase system of inclusions in a host matrix.  The inclusion phase consists of identical incompressible droplets randomly arranged in the solid elastic host matrix, as seen in Fig. \ref{fig:schematic}.  Under stress-free circumstances, the droplets are spherical. 

Benveniste \citep{Benveniste87} described the central assumption of the EIAS method as being equivalent to require that the fourth order tensor relating the average strain in a typical inclusion to the average strain in the matrix is equivalent to ``Wu's tensor'', $T_{ijkl}$' \cite{Wu66}.  Wu's tensor relates the uniform strain in an inclusion embedded in an ``all-matrix'' material to the imposed uniform strain at infinity. Here, the inclusion phase is denoted with the superscript *, and the matrix phase is free of indices. Across the interface between the phases in the equivalent inclusion system, the stress and displacement are continuous  (also known as perfect bonding conditions).  

For a composite consisting of spherical elastic inclusions with bulk/shear moduli $K^*,\mu^*$ embedded in a matrix with moduli $K,\mu$, the EIAS method gives that the effective composite moduli $\overline{K},\overline{\mu}$ are (see Eqs. (31) and (32) of \cite{Benveniste87}):
\be
\overline{K}=K+\phi(K^*-K)A_m, \quad \overline{\mu}=\mu+\phi(\mu^*-\mu)A_s\,,
\label{benveniste}
\ee
where
\begin{multline}
A_m=\frac{K}{K+(1-\phi)(K^*-K)S_m}, \quad A_s=\frac{\mu}{\mu+(1-\phi)(\mu^*-\mu)S_s},\\
\mathrm{and}\quad S_m=\frac{1+\nu}{3(1-\nu)}\quad S_s=\frac{8-10\nu}{15(1-\nu)}\quad\quad
\label{auxiliary}
\end{multline}
and $\nu$ is Poisson's ratio of the matrix.

Here we first show the equivalence between a droplet embedded in an elastic solid with an isotropic interfacial tension $\gamma$ and a corresponding elastic inclusion with no interfacial tension.
This allows us to calculate the moduli $K^*$ and $\mu^*$,  which we can then substitute into the above equations to yield the effective composite properties.

\section{Calculating the equivalent inclusion moduli}

Following Style et al., \cite{Stylesoft15, Style15}, we model the boundary condition for the elastic stress at the surface of the droplet using the Young-Laplace equation for the discontinuity of the traction vector $\sigma \cdot \textbf{n}$; 
\be
\sigma \cdot \textbf{n}=-p\textbf{n}+\gamma \mathcal{K} \textbf{n}, 
\ee
where $\textbf{n}$ is the normal to the deformed droplet surface, $p$ the pressure in the droplet, and the "total curvature" $\mathcal{K}$ is the sum of the local principal curvatures.  Importantly, the interfacial stress is treated as a constant, isotropic and strain-independent surface tension $\gamma$, which is an excellent approximation for a wide range of soft materials \cite[e.g.][]{Hui13}. 

The bulk modulus $K^*$ of the equivalent elastic inclusion can be calculated \cite[e.g.,][]{chp8micromechanics} by considering a spherical particle embedded in an infinite host material subjected to a spherically symmetric strain at infinity, yielding
\be
K^*=K_{incl}+\frac{2\gamma}{3 R} \,, 
\label{equivalentbulk}
\ee
where $R$ is the radius of the liquid inclusion, and $K_{incl}$ is the bulk modulus that the inclusion would have in absence of surface effects. Thus $K^*\rightarrow \infty$ as $K_{incl}\rightarrow\infty$ in our incompressible droplet inclusions.

We obtain the shear modulus $\mu^*$ of the equivalent elastic inclusion by comparing Eshelby's results for the elastic moduli of a dilute composite with spherical elastic inclusions \cite{Eshelbypg390}: 
\begin{subequations} 
\begin{align} 
&\overline{K}^{dil}=\frac{K}{1-\alpha^{-1} \phi},\quad \alpha=\frac{1+\nu}{3(1-\nu)}\,, \label{bulkdiluteapprox}\\
\overline{\mu}^{dil}=\frac{\mu}{1+B\phi}&, \quad B=\frac{\mu^*-\mu}{(\mu-\mu^*)\beta-\mu},\quad \beta=\frac{2}{15}\frac{4-5\nu}{1-\nu}\,\label{sheardiluteapprox} 
\end{align}
\end{subequations} 
to Style \emph{et al.}'s result for the Young's modulus of a dilute composite containing incompressible liquid droplets \cite[][Eq.\,(19)]{Stylesoft15}: 
\be
\frac{\overline{E}^{dil}}{E}=\left[ 1+\frac{3(1-\nu)\left[\frac{R}{L}(1+13\nu)-(9-2\nu+5\nu^2+16 \nu^3)\right]}{(1+\nu)\left[\frac{R}{L}(7-5\nu)+(17-2\nu-19 \nu^2)\right]}  \phi \right]^{-1}. 
\label{19Style}
\ee 
Using Eq. (\ref{bulkdiluteapprox}) and noting $\overline{\mu}^{dil}=3\overline{K}^{dil}\overline{E}^{dil}/(9\overline{K}^{dil}-\overline{E}^{dil})$, Eq. (\ref{19Style}) becomes
\be 
\frac{\overline{\mu}^{dil}}{\mu}=\frac{17-2\nu -19\nu^2 + \frac{R}{L}(7-5\nu)}{ 17 - 2 \nu - 19\nu^2 +15(\nu^2-1)\phi +\frac{R}{L}(7 -5\nu -15(\nu -1)\phi)}.
\label{shear_eqn}
\ee
Thus we equate Eqs. (\ref{sheardiluteapprox}) and (\ref{shear_eqn}) to obtain
\be
\frac{\mu^*}{\mu}=\frac{8(1+\nu)}{3(1+\nu)+5\frac{R}{L}}.
\label{equivalentshear}
\ee
This agrees with \cite{Ducloue14} in the limit of an incompressible matrix. Finally,  the equivalent Young's modulus is
\be \frac{E^*}{E}=\frac{3\mu^*}{2\mu(1+\nu)}=\frac{4}{1+\nu+\frac{5}{3}\frac{R}{L}}\,.
\label{equivalentyoung}
\ee 
It is important to note that although a finite value of the volume fraction $\phi\ll 1$ is assumed in the derivation of 
the equivalent moduli in Eqs  (\ref{equivalentbulk}), (\ref{equivalentshear}) and (\ref{equivalentyoung}), all are {\em independent of} $\phi$. 
In the case of an incompressible matrix ($\nu=1/2$) we recover the expression for $E^*$ of Style et al., \cite[][Eq.\,(9)]{Style15}, 
\be
\left(\frac{E^*}{E}\right)=\frac{24\frac{L}{R}}{10+9\frac{L}{R}}.
\ee
Now, for arbitrary Poisson's ratio $\nu$ of the host matrix, we find that (a) when $R\gg L$ the droplets behave like inclusions with Young's modulus $E^*=12\gamma/5R$, and (b) when $R \ll L$, in the capillarity-dominated regime, the equivalent Young's modulus of each inclusion saturates at $E^*=4E/(1+\nu)$. This shows that despite the widespread ansatz that $E^*=2\gamma/R$, the effective stiffness cannot become arbitrarily large as the droplet shrinks. Therefore, the limits (a) and (b) found by Style et al., \cite[][]{Stylesoft15} for an incompressible host matrix, $E^*\rightarrow 12\gamma/5R$  ($R\gg L$) and $E^*\rightarrow 8E/3$  ($R\ll L$) respectively, are consistent with a more general theory. 

\section{The effective composite moduli}

Having obtained equations for the the equivalent inclusion moduli, $K^*,\mu^*$, we can substitute them into Eqs. (\ref{benveniste}) and (\ref{auxiliary}) to obtain
\begin{widetext}
\be 
\frac{\overline{K}}{K}=\lim_{K^*\rightarrow \infty} \left[ 1 + \frac{\phi(K^*-K)}{K+(1-\phi)(K^*-K) S_m}\right]=1+\frac{\phi}{(1-\phi)S_m}\,
\label{effectivebulk}
\ee
and
\be
\frac{\overline{\mu}}{\mu}=1-\frac{15 (\nu-1) (1 -\frac{R}{L} +\nu)\phi}{\frac{R}{L} \left[7 + 8 \phi - 5 \nu (1 + 2 \phi)\right] +(1 +\nu)\left[17 -8 \phi +\nu (10 \phi -19)\right]}.
\label{effectiveshear}
\ee
Hence from Eq.\,(\ref{effectivebulk})  one finds
\be
\frac{\overline{E}}{E}= \frac{\frac{3}{2(1+\nu)(1-2\nu)}\left(\frac{\overline{K}}{K}\right)\left(\frac{\overline{\mu}}{\mu}\right)}{\frac{1}{1-2\nu}\frac{\overline{K}}{K}+\frac{1}{2(1+\nu)}\frac{\overline{\mu}}{\mu}}=\frac{\nu (4\phi-1)-(2\phi +1)}{1+\nu}\,
\frac{f_1+\frac{R}{L}f_2}{f_3+\frac{R}{L}f_4},
\label{effectiveYoung}
\ee
with
\be \left\{\begin{array}{l}f_1(\nu,\phi)= -(1 +\nu) \left[\nu (19 + 5 \phi)-(17+7 \phi)\right]\\
f_2(\nu,\phi)=  (5\nu-7) (\phi-1)\\
f_3(\nu,\phi)= (1 + \nu) (19 \nu-17) + \left[44\nu -14 +
    2 (5 - 24 \nu) \nu^2\right] \phi + \left[13 - 15 \nu + 
    2 \nu (15 \nu-13) + \nu (2 \nu-1) (-13 + 
       15 \nu)\right] \phi^2 \\
f_4(\nu,\phi)= 5 \nu-7 +2 (7 \nu-5) \phi + (1 - 2\nu) (15 \nu-13) \phi^2 \end{array}\right..
\ee
\end{widetext}
As expected in the dilute limit $\phi\rightarrow 0$ of Eq.(\ref{effectiveYoung})  we recover  Eq. (\ref{19Style}).

\begin{figure}[htp]
\centering 
\includegraphics[width=0.95\columnwidth]{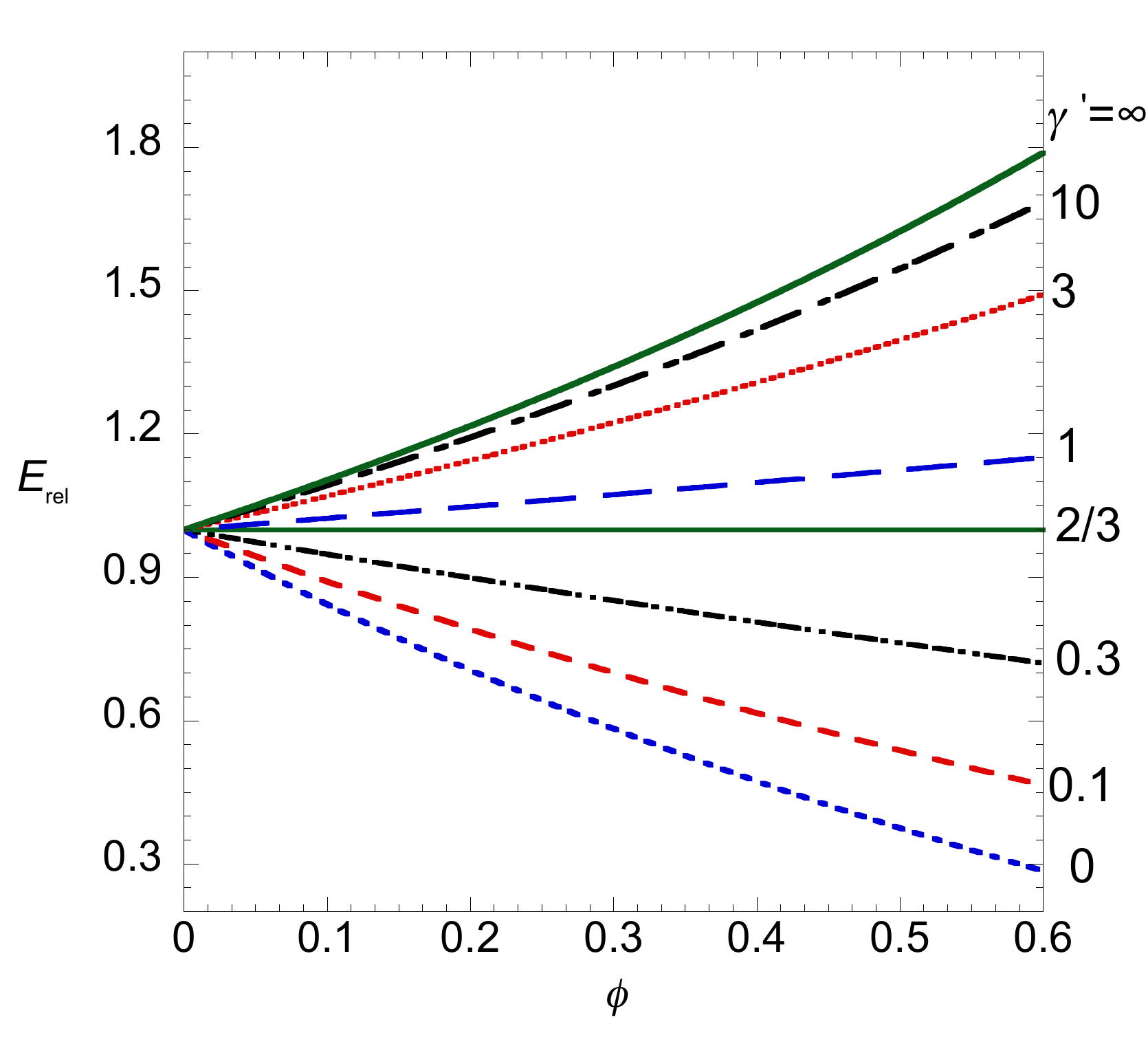}
%\subfigure[]{\includegraphics[width=0.85\columnwidth]{schematic}}
\vspace{-0.4cm}
\caption{In the incompressible matrix case $E_{rel}$ versus $\phi$ for a wide range of $\gamma^\prime$ from the softening to the stiffening regime, according to the EIAS theory.}
\label{fig:regular}
\end{figure}

\begin{figure}[htp]
\centering 
\includegraphics[width=0.95\columnwidth]{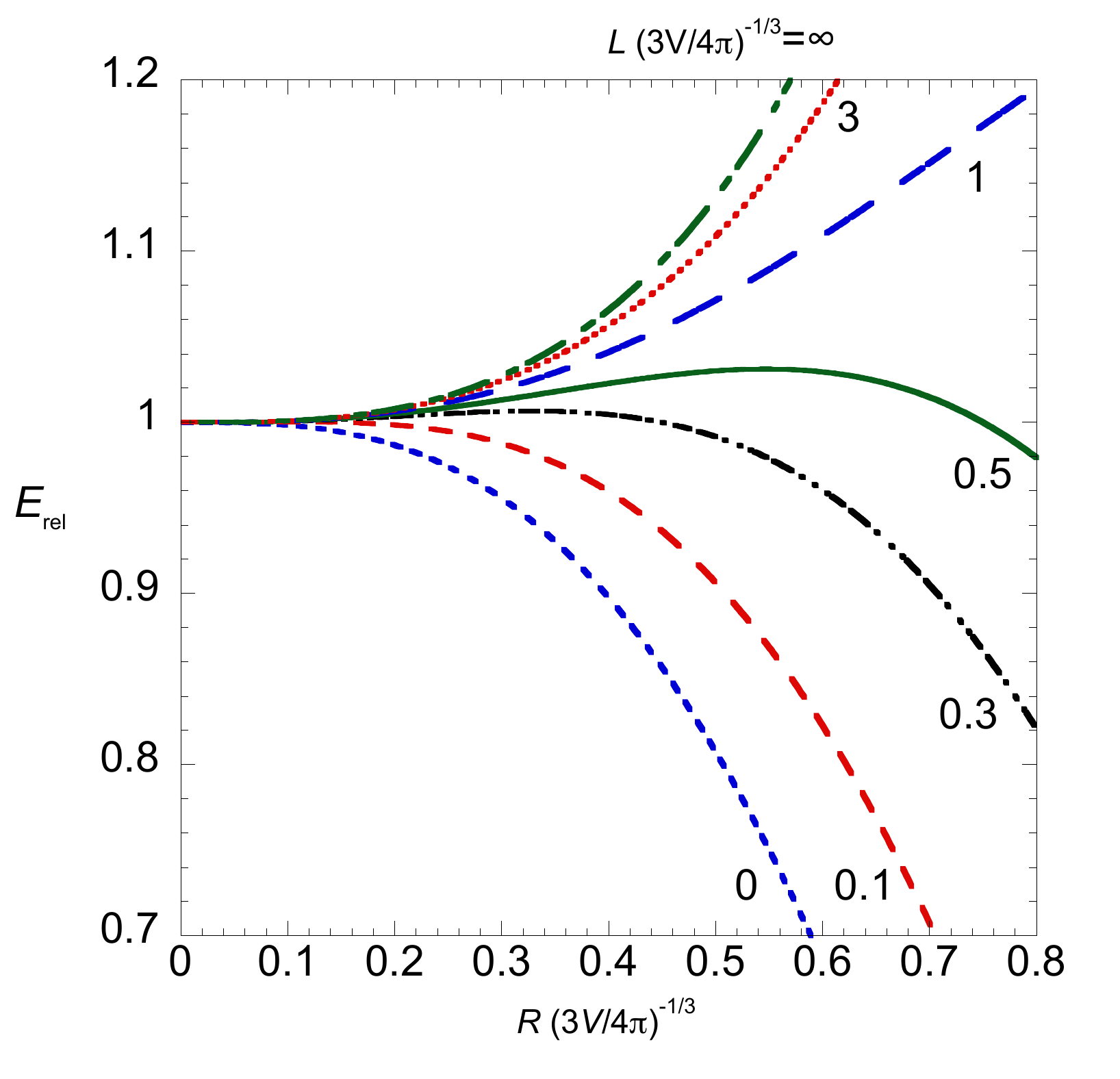}
%\subfigure[]{\includegraphics[width=0.85\columnwidth]{schematic}}
\vspace{-0.4cm}
\caption{In the incompressible matrix case $E_{rel}$ versus $R/[ (3V / 4 \pi )^{1/3}]$ for a wide range of the parameter $L/[ (3V / 4 \pi )^{1/3}]$, according to the EIAS theory.}
\label{fig:R}
\end{figure}

Next we focus on the special case of an incompressible matrix ($\nu=1/2$), where the identity $E_{rel}\equiv\left(\frac{\overline{E}}{E}\right)=\left(\frac{\overline{\mu}}{\mu}\right)\equiv\mu_{rel}$ holds. In some experimental situations it is interesting to know the effective Young's modulus for an incompressible matrix with a finite concentration of inclusions of arbitrary size where the bulk elasticity ($R\gg L$) and the capillarity dominated ($R\ll L$) limits manifest themselves.  Here, Eq.\,(\ref{effectiveYoung}) takes the simpler form
\be
\left(\frac{\overline{E}}{E}\right)=\frac{15+9\phi + \frac{R}{L} (6-6\phi)}{15-6\phi +\frac{R}{L}(6 + 4\phi)},
\label{moritanakaincompressible}
\ee
whose large and small droplet limits are  
\be
\left(\frac{\overline{E}}{E}\right)= \left\{\begin{array}{ll}\frac{3-3\phi}{3+2\phi}, & R \gg L \\
\frac{5+3\phi}{5-2\phi}, & R \ll L 
\end{array}\right..
\label{moritanakalargesmalldrop}
\ee
When, as above, the elastocapillary length is based on the matrix material, $L\equiv \gamma/E$, a natural dimensionless parameter is $\gamma^\prime\equiv L/R=\gamma/(E R)$, which we use to rewrite Eq. (\ref{moritanakaincompressible}) as 
\be
E_{rel}\vert_{\nu=1/2}=\frac{2-2 \phi + \gamma^\prime (5+ 3\phi)}{2+\frac{4}{3} \phi +\gamma^\prime(5 - 2\phi)}\;.
\label{moritanakaincompressibleintermsofgamma}
\ee
Fig. \ref{fig:regular} shows $E_{rel}$ of Eq. (\ref{moritanakaincompressibleintermsofgamma}) versus $\phi$, and in Fig. \ref{fig:R} it is plotted against $R/[ (3V / 4 \pi )^{1/3}]$, where $V$ is the volume of composite per inclusion.  We see in Fig. \ref{fig:regular} that the
$\gamma^\prime = L/R < 2/3$ ($\gamma^\prime > 2/3$) softening (stiffening) behavior spans the experimental range seen by Style et al., \cite{Style15}.  
We also find exact ``mechanical cloaking'', where $E_{rel}$ is constant at $\gamma^\prime = 2/3$ for all liquid volume fractions.  Precisely the same cloaking condition is found in the dilute theory \cite{Stylesoft15}, {\em and} from a complimentary generalized 3-phase self-consistent approach \cite{MSWb} (again independent of $\phi$).

As $\gamma^\prime$ becomes arbitrarily large (or the droplets become arbitrarily small), Eq.\,(\ref{moritanakalargesmalldrop}) shows that the capillary-dominated stiffening regime asymptotes to $E_{rel}\vert_{\nu=1/2} \rightarrow (5+3\phi)/(5-2\phi)$.  This is the upper limit of rigidity, scaling as $1/(1-\phi)$ at small $\phi$ (Fig. \ref{fig:regular}).    

\begin{figure}[htp]
%\vspace{-0.4cm}
\centering 
\includegraphics[width=1\columnwidth]{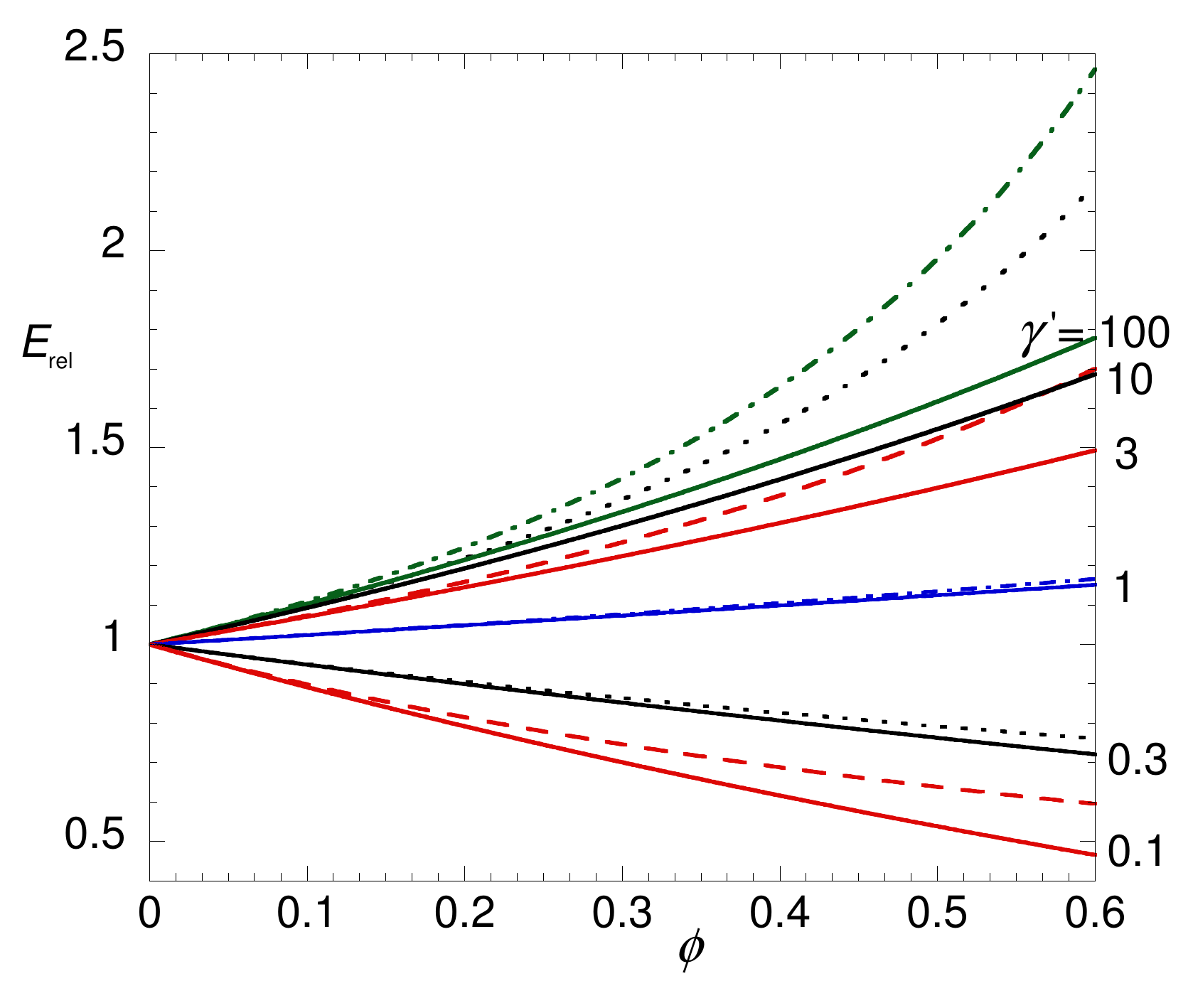}
%\subfigure[]{\includegraphics[width=0.85\columnwidth]{schematic}}
\caption{In the incompressible matrix case $E_{rel}$ versus $\phi$ for the EIAS (solid curves, Eq. \ref{moritanakaincompressibleintermsofgamma}) and dilute (dotted/dashed curves, \cite{Stylesoft15}) theories over a wide range of $\gamma^\prime$ from the softening to the stiffening regime. From the bottom to the top, the solid EIAS curves correspond to $\gamma^\prime=0.1$, $0.3$, $1$, $3$, $10$, $100$. The dilute theory curves correspond to $\gamma^\prime=0.1$ (red, dashed line), $0.3$ (black, dotted line), $1$ (blue, dash-dotted line), $3$ (red, dashed line), $10$ (black, dotted line), $100$ (green, dash-dotted line).}
\label{Mori}
\end{figure}

\begin{figure}[htp]
\vspace{-0.4cm}
\centering 
\includegraphics[width=1\columnwidth]{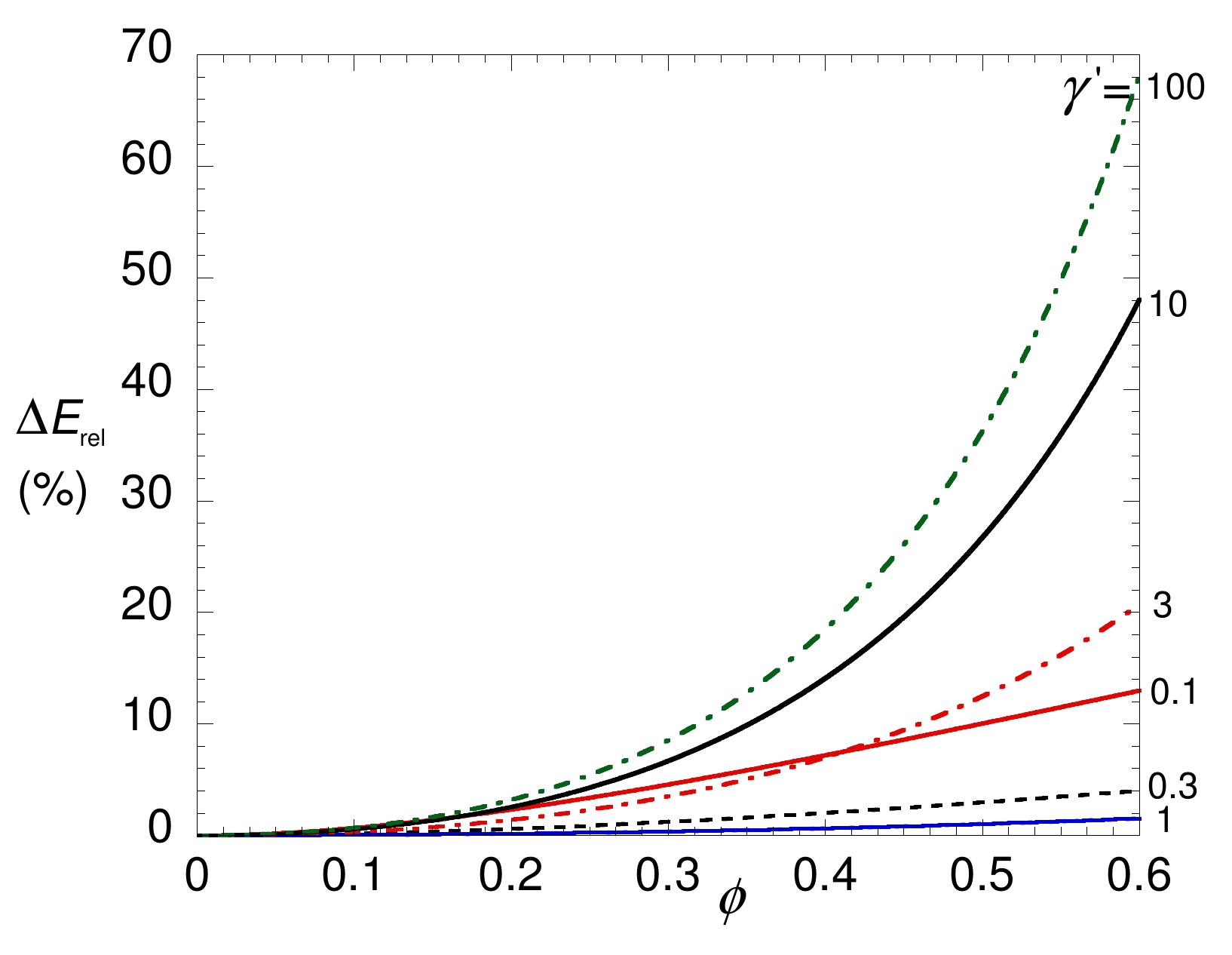}
%\subfigure[]{\includegraphics[width=0.85\columnwidth]{schematic}}
\caption{In the incompressible matrix case, the percent deviation between the dilute and EIAS theories $\Delta E_{rel}$ versus $\phi$, shown for the same values of $\gamma^\prime$ as in Fig.\,\ref{Mori}: $0.1$ (red, solid line), $0.3$ (black, dashed line), $1$ (blue, solid line), $3$ (red, dash-dotted line), $10$ (black, solid line), $100$ (green, dash-dotted line). }
\label{Morideviation}
\end{figure}

Finally, we note that in the limit $\phi\rightarrow 0$, the present theory quantitatively captures the dilute theory \cite{Stylesoft15}.
In Figs. \ref{Mori} and \ref{Morideviation}, we compare these two predictions of $E_{rel}$ for the incompressible matrix case. Clearly, the EIAS theory is softer than the dilute theory in both the softening ($\gamma^\prime < 2/3$) and the stiffening ($\gamma^\prime > 2/3$) regimes.
Importantly, for volume fractions up to  $\phi \approx 0.2$ (after which the dilute theory begins to break down), there is only a few percent deviation, as shown in Fig. \ref{Morideviation}.  This is well within experimental error \cite{Style15} and thus we show that the dilute theory provides an accurate and simple framework for comparison, given that it is the appropriate asymptotic limit of the non-dilute theory.  It is only when $\phi$ increases that large deviations appear, and these are most pronounced for large $\gamma^\prime$.   Independent of $\phi$, the two theories predict precisely the same mechanical cloaking condition of the inclusions; $\gamma^\prime=2/3$. 


\section{Conclusions} 
In light of recent work showing unexpected stiffening behavior of the effective elastic response of soft materials with liquid inclusions \cite{Stylesoft15, Style15}, we have revisited the Mori-Tanaka, or Equivalent Inclusion-Average Stress (EIAS), method for composite materials to account for the (strain-independent) liquid/matrix interfacial tension.   The motivation is that whilst Style et al., \cite{Stylesoft15, Style15} explained experimental data using a dilute theory, we sought to understand the limits of the dilute approximation by extending a known approach for non-dilute systems to account for the stiffening behavior associated with interfacial forces.  In so doing, we quantitatively analyzed when the dilute theory breaks down and thus confirmed that the comparison of experiment and theory \cite{Style15} occurred in the regime where the dilute theory is valid.  

In detail, we extended the EIAS theoretical framework for the effective elastic moduli of composites including liquid droplets, by taking into account the surface tension at the droplet host-matrix interface when the matrix is a linear-elastic material.  The dilute limit of the EIAS theory is achieved by taking $\phi \rightarrow 0$, and we find that the effective Young's modulus depends solely on two only parameters; $\phi$ and $\gamma^\prime=L/R$.  We examined this graphically only in the incompressible case  of $\nu=1/2$. These models, along with a generalized self-consistent 3-phase theory \cite{MSWb}, predict the same exact cloaking condition of the far-field signatures associated to the presence of the inclusions, viz., $R=3L/2$, independent of volume fraction $\phi$.  

There are a range of possible comparisons and tests that immediately come to mind.  For example, in situations wherein the host matrix is a nonlinear elastic \cite[e.g.,][]{Castaneda98,Jiang04}, or viscoelastic \cite[e.g.,][]{Palierne90} material. Finally, it would be of interest to compare this framework and that in our companion paper \cite{MSWb}, in which we treat the inclusion/matrix interface using a strain-independent surface tension, with approaches using an interfacial stress model \cite[e.g.,][]{Duan05,Duan07}. 

\section{Acknowledgments}\label{Acknowledgments}
FM and JSW acknowledge Swedish Research Council Grant No. 638-2013-9243 and the 2015 Geophysical Fluid Dynamics Summer Study Program at the Woods Hole Oceanographic Institution, which is supported by the National Science Foundation and the Office of Naval Research under OCE-1332750.  JSW also acknowledges 
a Royal Society Wolfson Research Merit Award.

\end{document}